\documentclass[12pt]{article}

\usepackage[mathscr]{eucal}
\usepackage{amssymb,latexsym}
\usepackage{verbatim}
\usepackage{graphicx}
\usepackage{amsmath}
\usepackage{amsthm}
\usepackage{enumerate}
\usepackage{authblk}
\usepackage{color}
\usepackage{url}

\usepackage{caption}

\usepackage{tikz}
\usetikzlibrary{matrix}
\usepackage{graphicx}
\usepackage{blindtext}

\usetikzlibrary{arrows,calc,shapes,decorations.pathreplacing}

\usepackage[normalem]{ulem}

\newtheorem{thm}{Theorem}[section]
\newtheorem{lem}[thm]{Lemma}
\newtheorem{Def}{Definition}[section]
\newtheorem{prop}[thm]{Proposition}
\newtheorem{rem}[Def]{Remark}

\title{Causal structure of evaporating black holes}

\author{Martin Lesourd}

\affil{University of Oxford, UK}

\begin{document}
\date{}
\maketitle
\vspace{.2in}

\begin{abstract}
We offer a mathematically rigorous basis for the widely held suspicion that full black hole evaporation is in tension with predictability. Based on conditions expressing the global causal structure of evaporating black hole spacetimes, we prove two theorems in Lorentzian geometry showing that such spacetimes either fail to be causally simple or fail to be causally continuous. These theorems, when combined with recent results \cite{AM} on the causal structure of spacetimes with timelike boundary, bear significantly on the question of whether these spacetimes permit for a predictable evolution. 
\end{abstract}



\section{Introduction}
Hawking \cite{H1}, \cite{H2}, \cite{H3} famously argued that quantum field theoretic effects are capable of disturbing a black hole to the extent of causing its gradual disappearance. His arguments immediately led many to suspect that spacetimes containing fully evaporated black holes cannot be fully predictable, classically or indeed semi-classically. Five years later, by way of a theorem in causal structure, Kodama \cite{K} attempted to formulate a mathematically rigorous basis for this suspicion. His theorem was then rewired and endorsed in an article by Wald \cite{Wald}, though Wald attributes the specific statement to Geroch.
\begin{thm}[Kodama, Geroch, and Wald \cite{K}, \cite{Wald}]
Let \((M,g)\) be a spacetime containing achronal subsets \(S_1, S_2\) where \(S_1,S_2\) is edgeless and \(S_1\) is connected. Then \(S_2\not\subset D^+(S_1)\) if the following holds:
\begin{enumerate} 
\item[(i)]  \(J^+(K)\cap S_2\) has compact closure, where \(K=S_1-D^-(S_2)\),
\item[(ii)] there is a point \(p\in D^+(S_1)-J^+(S_2)\cup J^-(S_2)\).
\end{enumerate}
\end{thm} 
The original motivation for this theorem was to apply to it to evaporating black hole spacetimes, where \(S_1\) and \(S_2\) are imagined to lie, respectively, before and after the black hole has evaporated, and \(p\) is a point in the black hole interior which is causally inaccessible from \(S_2\). In the asymptotically flat case, the spacetimes envisaged are as follows.
\begin{center}
\includegraphics[width=.6\linewidth]{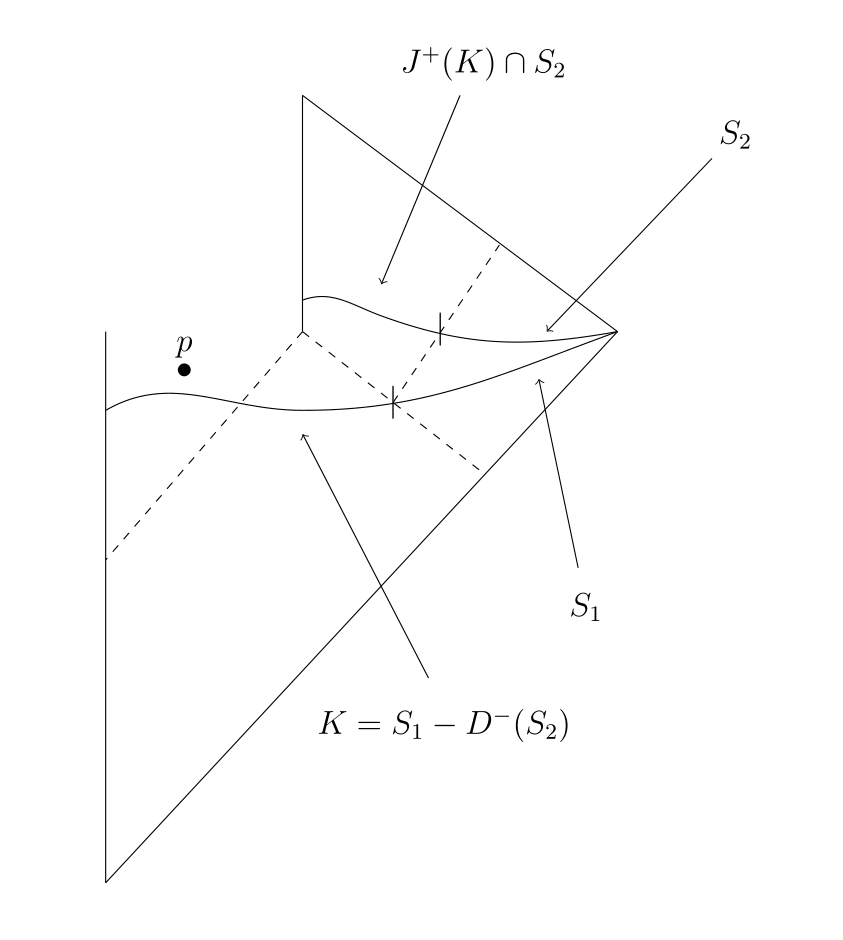}
\end{center}
\indent Theorem 1.1 appears to be the only mathematically rigorous study of causal structure in the context of evaporating black hole spacetimes. It is to be regarded, for this reason, as an important result in a topic where mathematically rigorous results are few and far between. With that being said, the theorem does not, in fact, ground the idea that evaporating black hole spacetimes are not predictable. In particular, the theorem shows that \(S_1\) cannot be a Cauchy surface for \(M\), but it does not actually rule out the possibility that \((M,g)\) is globally hyperbolic. Indeed, it is straightforward to construct globally hyperbolic spacetimes satisfying the conditions of theorem 1.1. We now describe one such example, which shows that \(S_2\not\subset D^+(S_1)\) could simply be due to a poor choice of \(S_1\) and \(S_2\).  \\ \\ 
\textbf{Example}. Take two dimensional Minkowski spacetime \((\mathbb{R}^2,\eta)\) in Cartesian co-ordinates. Remove everything but an open globally hyperbolic diamond with vertices given by \((x=\pm 1,t=0)\) and \((x=0,t=\pm 1)\). Remove from this diamond spacetime a further diamond with vertices given by \((x=0,t=-1)\), \((x=\pm \frac{1}{2},t=-\frac{1}{2})\), \((x=0,t=0)\). The new spacetime is still globally hyperbolic but the conditions of theorem 1.1 are easily satisfied by suitably choosing \(S_1\) and \(S_2\); consider, for instance, the following diagram.
\begin{center}
\includegraphics[width=.7\linewidth]{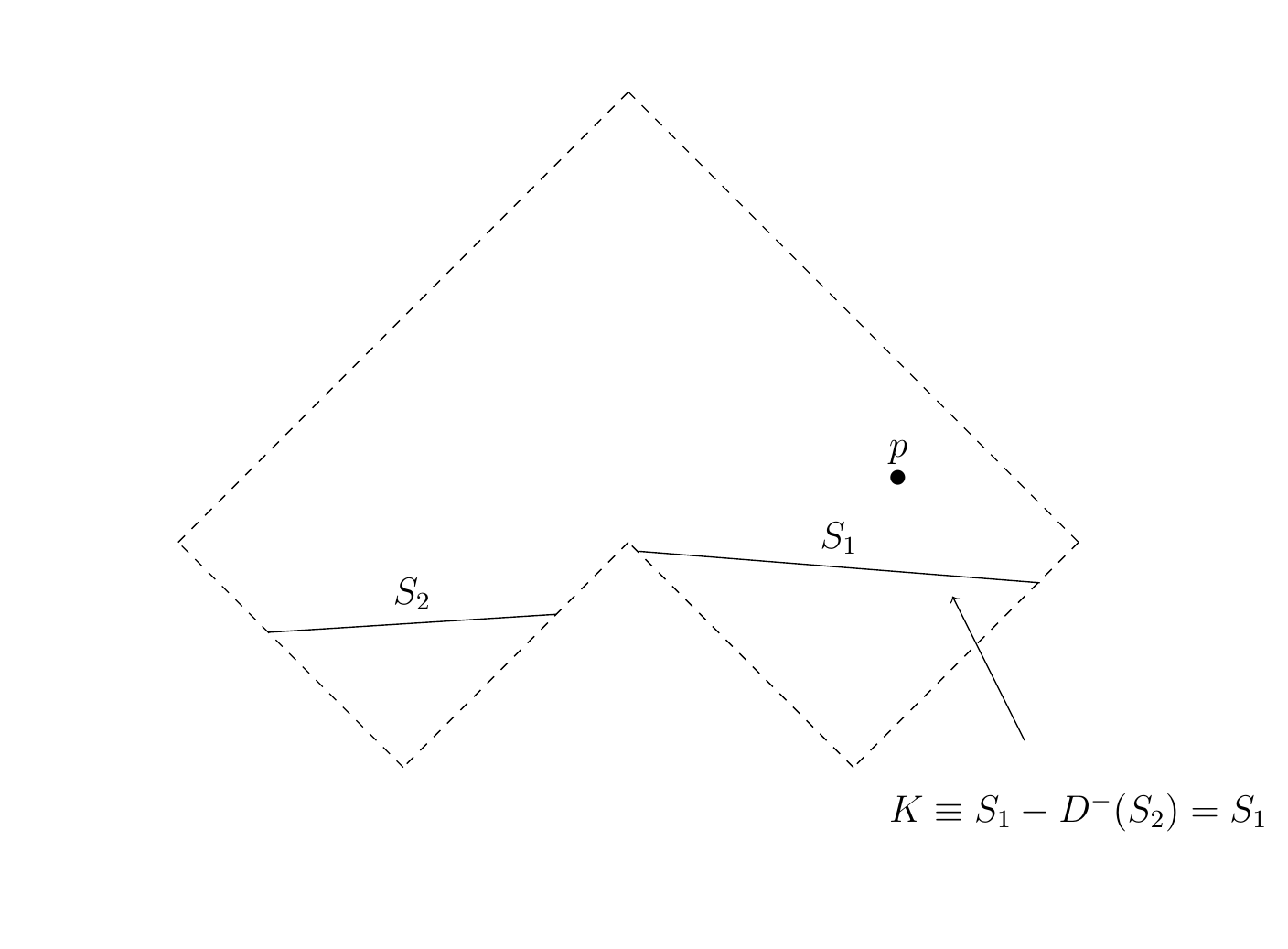}
\end{center}

\indent In view of such examples (and less trivial ones one can construct), we cannot rely on theorem 1.1 alone as a basis for the widely discussed suspicion that fully evaporating black holes lead to a failure in predictability. The purpose of this article is to offer one such mathematically rigorous basis for this suspicion. More specifically, we shall prove two theorems to the effect that spacetimes containing evaporating black holes either fail to be causally simple or causally continuous.\footnote{Recall that causal simplicity and continuity sit second and third from top in the causal hierarchy of spacetimes: Global Hyperbolicity \(\Rightarrow\) Causal Simplicity \(\Rightarrow\) Causal Continuity \(\Rightarrow\) Stably Causal \(\Rightarrow\) Strongly Causal \(\Rightarrow\) Distinguishing \(\Rightarrow\) Causal \(\Rightarrow\) Chronological \(\Rightarrow\) Non-totally Vicious.} \\ \indent One of our theorems proves the failure of causal simplicity and it may be thought of as an extension of theorem 1.1; that is, we add conditions which rules out various examples (as above) that one may construct. \\ \indent Our other theorem is formulated entirely differently and assumes, from the outset, the existence of an event horizon. Its conclusion is stronger however, as \((M,g)\) is shown to be causally discontinuous. For simplicity, we define the event horizon using Penrose's concept of aymptotic flatness at null infinity. This can be generalized to spacetimes with more general asymptotics, and indeed we plan to do so in future work.   \\ \indent
In section 2 we recall some preliminaries of Lorentzian geometry. Sections 3 and 4 are dedicated, respectively, to the main theorems. Finally, in section 5 we discuss possible improvements and certain open questions. \\  \\
\textbf{Acknowledgements} I thank the Ruth and Nevil Mott Scholarship and the AHRC for funding this research. I also thank Erik Curiel, John Manchak, Claudio Paganini, Rafael Sorkin, Andrew Strominger, Chris Timpson, Bob Wald and James Weatherall for useful exchanges. I also thank, dearly, Carina Prunkl for producing the illustration herein. Finally, I thank John Earman, for insightful remarks and for his book - ‘Bangs, Crunches, Whimpers, and Shrieks: Singularities and Acausalities in Relativistic Spacetimes’ - which played a part in inspiring this work. Finally, I thank two anynomous reviewers, whose comments contributed to a greatly improved version of this article.

\section{Preliminaries in Lorentzian geometry}
Here we list some standard results that will be used below. Chapter 3 of \cite{B} provides the relevant background for the definitions.
\begin{Def}
A spacetime \((M,g)\) is non-total imprisoning when no future or past inextendible causal curve is contained in a compact set. 
\end{Def}
\begin{Def}
A spacetime \((M,g)\) is said to be past reflecting if \(I^+(p) \subset I^+(q)\Rightarrow I^-(q)\subset I^-(p)\). Future reflecting is defined dually. A spacetime is reflecting if it is both future and past reflecting.
\end{Def}
\begin{Def}[\cite{B}]
The set-valued function \(I^+\) is inner continuous at \(p\in M\) if, for each compact set \(K\subseteq I^+(p)\), there exists a neighborhood \(U(p)\) of \(p\) such that \(K \subseteq I^+(q)\) for each \(q\in U(p)\). 
\end{Def}
\begin{Def}[\cite{B}]
The set-valued function \(I^+\) is outer continuous at \(p\in M\) if, for each compact set \(K\subseteq M\backslash \overline{I^+(p)}\), there exists some neighborhood \(U(p)\) of \(p\) such that \(K\subseteq M\backslash \overline{I^+(q)}\).
\end{Def}
\begin{lem}[Hawking-Sachs \cite{HS}]
For any spacetime \((M,g)\) the outer continuity of \(I^+\) and \(I^-\) is equivalent to the reflecting property.
\end{lem}
\begin{Def}[\cite{B}]
A spacetime \((M,g)\) is said to be causally continuous if and only if it is distinguishing and the set-valued functions \(I^+\) and \(I^-\) are outer continuous.
\end{Def}
\begin{Def}
A spacetime \((M,g)\) is said to be causally simply if it is causal and \(J^{+(-)}(p)\) is closed for all \(p\in M\). 
\end{Def}
\begin{Def}[\cite{Galloway}]
Let \(S \subset M\) be achronal. Then \(p \in S\) is an edge point of \(S\) provided every neighborhood \(U(p)\) of \(p\) contains a timelike curve \(\gamma\) from \(I^-(p,U)\) to \(I^+(p,U)\) that does not meet \(S\). We denote by \(edge(S)\) the set of edge points of \(S\).
\end{Def}

\begin{prop}[\cite{Galloway}]
Let \(S\) be closed. Then each \(p \in \partial I^+(S)\backslash S\) lies on a null geodesic contained in \(\partial I^+(S)\backslash S\), which either has a past endpoint on \(S\), or else is past inextendible in \(M\).
\end{prop}
\begin{prop}[\cite{Galloway}]
Let \(S\) be an achronal subset of a spacetime \(M\). Then \(H^+(S) \backslash edge(S)\), if nonempty, is an achronal \(C^0\) hypersurface of \(M\) ruled by null geodesics, each of which is either past inextendible in \(M\) or has a past endpoint on \(edge(S)\).
\end{prop}
The Lorentzian distance function \(d(.,.)\) and the Lorentzian length \(l_\gamma(p,q)\) is defined as in \cite{Ming}, from which we shall also use the following definition. 
\begin{Def}[Definition 2.11 of \cite{Ming}]
A continuous causal curve \(\gamma:I\to M\) is maximizing if, for every \(t_1,t_2\subset I\), \(t_1<t_2\), \(d(\gamma(t_1),\gamma(t_2))=l(\gamma\mid_{[t_1,t_2]})\). A sequence of continuous causal curves \(\gamma_n:I_n\to M\), is limit maximizing if defined \[\epsilon_n=\sup_{t_1,t_2\in I_n,t_1<t_2}[d(\gamma_n(t_1),\gamma_n(t_2))-l(\gamma_n\mid_{[t_1,t_2]})]\geq 0\] it is \(\lim_{n\to \infty}\epsilon_n=0\).
\end{Def}

\section{An extension of theorem 1.1} 
In theorem 1.1, \(S_1,S_2\) are achronal, edgeless (and thus closed by \cite{ON} p. 414, corollary 26) and \(S_1\) is connected. Below, we make the additional assumption that \(S_1\) is non-compact and that \(S_2\) is connected. 
\begin{thm}
Let \((M,g)\) be a non-totally imprisoning spacetime with connected, achronal, edgeless sets \(S_1,S_2\subset I^+(S_1)\) such that \(S_1\) is closed and non-compact. Suppose the following: 
\begin{enumerate}
\item[(i)] \(J^+(K)\cap S_2\) has compact closure, where \(K=S_1-D^-(S_2)\), 
\item[(ii)] there is a point \(p\in D^+(S_1) - J^+(S_2)\cup J^-(S_2)\), 
\item[(iii)] \(S_1 - V \subset J^-(S_2)\), where \(V\) is a compact subset of \(S_1\), 
\item[(iv)] \(J^-(Q)\cap S_1\) has compact closure where \(Q=S_2-D^+(S_1)\).
\end{enumerate}
Then \((M,g)\) is not causally simple.
\end{thm}
\textbf{Remark}. Condition (iii) blocks the kind of scenario described in our example of section 1, and it also expresses the idea that \(S_1\) and \(S_2\) are related as in figure 1, so that, in particular \(V\) may represent the portion of \(S_1\) which is inside the black hole. With figure 1 in mind, condition (iv) expresses the idea that the non-compact ends of \(S_2\) are contained in \(D^+(S_1)\), so that in effect we are assuming a form of predictability `far away' from the black hole.
\begin{proof}
We shall assume causal simplicity and obtain a contradiction. \\ \indent 
First we apply theorem 1.1 to deduce that \(H^+(S_1)\neq \emptyset\). Since \(S_1\) is closed and edgeless, we apply proposition 2.3 to infer that \(H^+(S_1)\) is ruled by past inextendible null geodesics not intersecting \(S_1\). \\ \indent
Second, we observe that \(H^+(S_1)\cap S_2\neq \emptyset\). Since \(S_2\not\subset D^+(S_1)\) and \(S_2\) is connected, \(H^+(S_1)\cap S_2\neq \emptyset\) will follow if we can show that \(S_2\cap D^+(S_1)\neq \emptyset\). Seeking a contradiction, we suppose that \(D^+(S_1)\cap S_2 = \emptyset\). In that case, by defining \(Q\) to be equal to \(S_2-D^+(S_1)\), we have \(Q=S_2\). Condition (iv) now becomes that \(J^-(S_2)\cap S_1\) has compact closure. But this is in contradiction with (iii). In particular, since \(S_1\) is closed and non-compact, the assumption that \(S_1-V \subset J^-(S_2)\) where \(V\) is a compact subset of \(S_1\) implies that \(J^-(S_2)\cap S_1\) has non-compact closure.  \\ \indent 
These remarks now permit us to consider a point \(x\in H^+(S_1)\cap S_2\). Since \(J^-(x)\) is closed, every point \(y\in \partial J^-(x)\) is also in \(J^-(x)\). By proposition 2.2, it then follows that for any such point \(y\), there exists a past directed achronal null geodesic with endpoints \(y\) and \(x\). These geodesics generate the boundary of the past lightcone of \(x\). \\ \indent By standard properties of Cauchy horizons, \(H^+(S_1)\) is an achronal hypersurface that is generated by null geodesics. By restricting attention to a sufficiently small convex neighborhood \(U(x)\) of \(x\), it follows that there is a unique achronal null geodesic \(\eta\) which belongs to both \(U(x)\cap H^+(S_1)\) and \(U(x)\cap \partial J^-(x)\).  \\ \indent 
Since \(U(x)\) is a convex neighborhood, we may take \(\eta\) to have two endpoints, \(x\) and \(z\), where \(z\in H^+(S_1)\cap U(x)\cap \partial J^-(x)\). Since \(S_1\) is edgeless, if \(\eta\) is extended towards the past as an unbroken achronal null geodesic, then it will eventually become past inextendible, it will not intersect \(S_1\), and it will remain entirely on \(H^+(S_1)\). \\ \indent The sought after contradiction will be to show that (iv) and the assumption of causal simplicity actually permit us to construct an achronal extension of \(\eta\) with past endpoint on \(S_1\).  \\ \indent 
Consider the set \(I^-(x)\cap S_1\). Clearly, this set is both non-empty and open. By assumption (iv) and by the standard identity \(\partial I=\partial J\) - eg., see page 191, chapter 8 of \cite{Wald} - it follows that \(\overline{I^-(x)\cap S_1}\) is compact. Thus, any infinite sequence of points in \(\overline{I^-(x)\cap S_1}\) has a limit point in \(\overline{I^-(x)\cap S_1}\). We may then consider an infinite sequence of points \(\{q_i\}\in I^-(x)\cap S_1\) with limit point \(q \in \partial (I^-(x)\cap S_1)\). For each \(q_i\), we consider a future directed timelike curve with past endpoint \(q_i\) and future endpoint \(r_i\in I^-(x)\) where \(r_i\to x\). We label each such curve by \(\{\delta_i\}\), each of which are defined on \([0,t_i]\), so that \(\delta_i(0)=q_i\) and \(\delta_i(t_i)=r_i\) for some \(0<t_i\).\\ \indent Since \(q\) is an accumulation point of a sequence of continuous timelike curves \(\{\delta_i\}\), part (1) of theorem 3.1 in \cite{Ming} permits us to extract a subsequence, \(\{\delta_k\}\). Clearly, \(q_k<< r_k\) for all \(k\) and the curves \(\delta_{k}\) have endpoints \(r_k,q_k\in I^-(x)\). We now show that causal simplicity implies that we can construct our sequence of curves, \(\{\delta_i\}\), such that our subsequence \(\{\delta_k\}\) converges to a smooth achronal limit curve \(\delta\) with endpoints \(q,x\). Since each distinct point on \(\partial (I^-(x)\cap S_1)\) gives rise to a distinct limit curve, it follows that one of these coincides with \(\eta\). Thus, the proof will complete for we will have produced an achronal extension of \(\eta\) with endpoint on \(\partial (I^-(x)\cap S_1)\). \\ \indent 
To show that \(\delta\) can be suitably constructed, we consider the following result of Minguzzi. 
\begin{thm}
A non-totally imprisoning spacetime \((M,\{g\})\) is causally simple if and only if, for every metric \(g\) in the conformal class \(\{g\}\), the Lorentzian distance function \(d_g(.,.)\) is continuous on the vanishing distance set. 
\end{thm}
Theorem 3.2 implies that we may take the distance function \(d_g(.,.)\) to be continuous on the vanishing distance set. This in turn implies that, for any infinite sequence of points \(\{p_n\}\), \(\{s_n\}\) with \(s_n\in I^+(p_n)\) and \(p\in \partial I^+(s)\) where \(p_n\to p\), \(s_n\to s\), we have \(d(p_n,s_n)<\epsilon\) as \(n\to \infty\). In our case, given that \(q\in \partial I^-(x)\), we have \(d(q,x)=0\). Continuity of \(d(.,.)\) implies that \(d(q_k,r_k)< \epsilon_k\), and thus that \(d(q_k,r_k)\to 0\) as \(k\to \infty\). By definition 2.5, this implies that may take the sequence \(\{\delta_k\}\) to be limit maximizing. By part (1) of theorem 3.1 in \cite{Ming}, the subsequence \(\{\delta_k\}\) converges \(h\)-uniformly on compact subsets to a maximizing limit curve \(\delta\). Seeing as \(q\in \partial I^-(x)\), \(\delta\) must in fact be an achronal null geodesic with past, future endpoint \(q\), \(x\). 
\end{proof}
\begin{rem}
It is likely that recourse to theorem 3.2 is unnecessary for the proof of theorem 3.1. Its inclusion here has the advantage of making the current argument particularly explicit. 
\end{rem}

\section{An alternative approach}
We now consider an alternative approach which has the virtue of proving a stronger statement than theorem 3.1. The price to be paid for this stronger conclusion is the posit of an event horizon along with a black hole region obeying certain specific properties, though in fact these properties are standard in the usual non-evaporating context and in any case the existence of an event horizon is an important part of the arguments in favor of evaporation. \\ \indent To formulate the theorem below, we shall specificy asymptotic boundary conditions. For simplicity, we have chosen Penrose's \cite{Pen} definition of \textit{asymptotic flatness at null infinity}, which we now recall. \\ \indent
Let \((M,g)\) be a four dimensional, chronological, connected spacetime which can be conformally included into a spacetime-with-boundary \((M',g')\) such that \(M\) is the interior of \(M'\), \(M=M'\backslash M'\). Assume, with regards to the conformal factor, that there exists a smooth function \(\Omega\) on \(M'\) such that: 
\begin{enumerate} 
\item[(i)] \(\Omega>0\) and \(g'=\Omega^2 g\) on \(M\),
\item[(ii)] \(\Omega=0\) and \(d\Omega \neq 0\) along \(M'\). 
\end{enumerate}
The boundary \(\mathcal{J}\equiv \partial M'\) is assumed to consist of two components, \(\mathcal{J}^+\) and \(\mathcal{J}^-\), which are smooth null hypersurfaces representing, respectively, future and past null infinity. A spacetime \((M,g)\) satisfying the above is said to be \textit{asymptotically flat at null infinity}.\\ \indent
We now prove the following theorem.
\begin{thm}
Let \((M,g)\) be a spacetime which is asymptotically flat at null infinity, such that its conformal boundary consists of two disconnected null components \(\mathcal{J} \equiv \mathcal{J}^+\cup \mathcal{J}^-\) each having topology \(V\times \mathbb{R}\). Suppose that the following properties obtain: 
\begin{enumerate}
\item[(i)] there is a non-empty event horizon \(\partial I^-(\mathcal{J}^+)\) and a non-empty black hole region defined by \(B\equiv I^+(\partial I^-(\mathcal{J}^+))\) such that \(\partial B = \partial I^-(\mathcal{J}^+)\) and \(B\cap I^-(\mathcal{J}^+)=\emptyset\),
\item[(ii)] \(\partial I^-(\mathcal{J}^+)\subset \overline{I^-(\Sigma)}\) where \(\Sigma\) is a complete cross section of \(\mathcal{J}^+\), i.e., a spacelike embedded submanifold of \(\mathcal{J}^+\) with topology \(V\).
\end{enumerate} 
Then \((M,g)\) is causally discontinuous. 
\end{thm}
\textbf{Remark 4.2}. That the event horizon is the boundary of a black hole region from which there is no escape to \(\mathcal{J}^+\), and that \(\mathcal{J}^+\) has topology \(\mathbb{R}\times V\), are both conditions which are standard in the usual setting.\footnote{Note that we need not assume that \(V\) has topology \(S^2\).} So far, then, the conditions are all standard. \\ \indent 
The main new conditions capturing the notion of evaporation is (ii), which demands that the event horizon is entirely contained in the closure of the past of some cross section of \(\mathcal{J}^+\). Here, we are imagining that the event horizon eventually ceases to exist, and, thus, that causal curves persisting sufficiently far into the future can \textit{almost} see the event horizon. Condition (iii) expresses what is meant by `sufficiently far', where moreover the properties of the mentioned generator follow from standard properties of \(\mathcal{J}^+\). The following diagram provides a means of visualizing the kinds of spacetime for which theorem 4.1 applies. 
\begin{center}
\includegraphics[width=.7\linewidth]{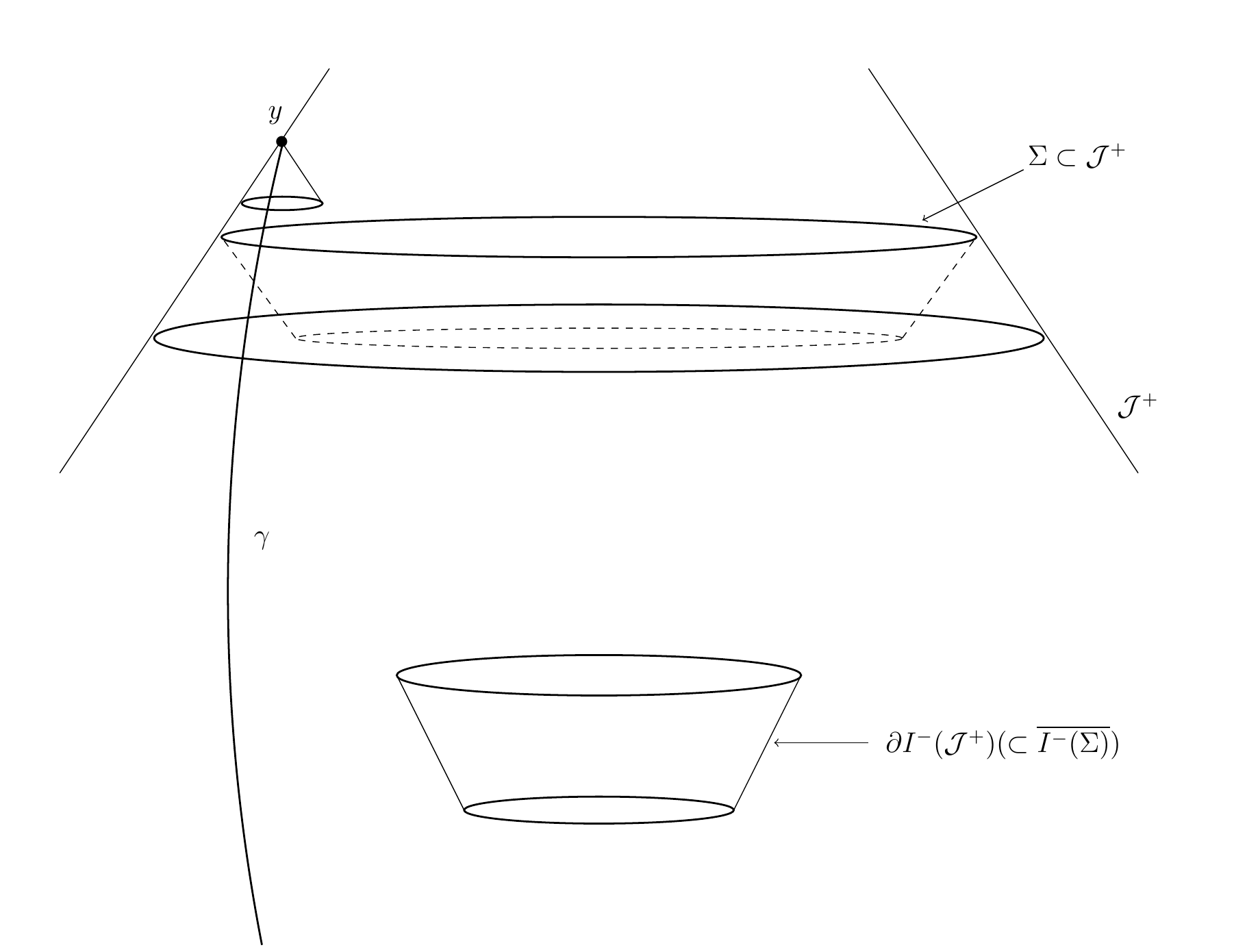}
\end{center}

\begin{proof}
We assume both outer continuity and reflectivity and derive a contradiction. \\ \indent 
Before going further, we note that by the above assumptions, there exists a non-trivial causal curve \(\gamma\) with future endpoint \(y\in \mathcal{J}^+\) where \(y\) is the future endpoint to a non-trivial achronal null geodesic generator of \(\mathcal{J}^+\) with past endpoint on \(\Sigma\); consider the figure above. \\ \indent 
First we make a preliminary observation. For any point \(x\in \partial I^-(\mathcal{J}^+)\), we have by assumption that \(x\in I^-(B)\cap \partial B \). Let \(U(x)\) be a small open neighborhood of \(x\). Consider the spacetime given by \((U(x),g')\) where \(g'\) is the spacetime metric of \(M\) restricted to \(U(x)\). In this spacetime, \(\partial I^-(\mathcal{J}^+)\cap U(x)\) is an achronal boundary without edge in \((U(x),g')\). Denote the achronal boundary \(\partial I^-(\mathcal{J}^+)\cap U(x)\) in \((U(x),g')\) by \(V\). By proposition 3.15 in \cite{Penrose} on achronal boundaries, \(V\) separates the spacetime \((U(x),g')\) such that \(U(x)= I^+(V)\cup V\cup I^-(V)\). It then follows that any point in \(I^-(V)\) is in \(I^-(B)\), that any point in \(I^+(V)\) is in \(B\) and that any point on \(V\) is in \(\partial I^-(\mathcal{J}^+)\). \\ \indent 
By assumption, \(\gamma\) has endpoint \(y \in \mathcal{J}^+\), and there is a point on the event horizon which is also in \( \partial I^-(y)\cap M\). Denote such a point by \(x\). Since \(\mathcal{J}^+\) is open, we may consider an open neighborhood \(U(y)\) of \(y\) such that \(U(y) \subset \mathcal{J}^+\). In this neighborhood, we may consider a point \(y'\in U(y)\) such that \(I^-(y) \subsetneq I^-(y')\). Note that by condition (i), it follows that \(I^-(y') \cap B=\emptyset\), and thus \(I^+(x)\cap I^-(y') =\emptyset\). \\ \indent 
Now consider a compact set \(K\subset M\) with non-empty interior such that \(K \subset I^-(y') - [\overline{I^-(y)}\cup \overline{I^-(\Sigma)}]\). We know that such a set exists because (ii) implies that the region \(I^-(y')-[\overline{I^-(y)} \cup \overline{I^-(\Sigma)}]\) is non-empty. Moreover, we know that such a set lies in an open region because \(I^-\) is open and \(\overline{I^-}\) is closed. \\ \indent 
By suitably choosing a compact set \(K\) and assuming reflectivity, we shall obtain a contradiction with outer continuity. \\ \indent First, we show that \(K\) can be chosen such that \(K\subset M-\overline{I^+(x)}\). Since we have chosen \(K\) such that \(K\subset I^-(\mathcal{J}^+)\), it follows that \(K\not\subset {I^+(x)}\). This means that, if \(K\) intersects \(\overline{I^+(x)}\), then it does so in a closed and achronal set, which we write as \(Z\equiv K\cap \partial I^+(x)\). We note that we are here choosing \(K\) such that \(K \not\subset Z\). This is clearly possible for \(K\) was originally chosen in the open region \(I^-(y')-[\overline{I^-(y)} \cup \overline{I^-(\Sigma)}]\). \\ \indent 
Let \(z\) be any point in \(Z\). By the condition that \(K \cap \overline{I^-(\Sigma)} =\emptyset\) and that \(\overline{I^-(\Sigma)}\) contains \(\partial I^-(\mathcal{J}^+)\), it is clear that \(z\notin \partial I^-(\mathcal{J}^+)\). Thus, \(z\) lies neither on the event horizon nor inside the black hole region \(B=I^+(\partial I^-(\mathcal{J}^+))\). It follows, then, that \(z\) is in \(I^-(\mathcal{J}^+)\). Since this is true for all points in \(Z\), we conclude that \(Z\), which is an achronal set in \(\partial I^+(x)\), is in \(I^-(\mathcal{J}^+)\). Now, \(Z\) is either empty or not. If it is then of course \(K\subset M-\overline{I^+(x)}\). If \(Z\neq \emptyset\), we can show that we can amend our choice of \(K\) to obtain a compact set which is in \(M-\overline{I^+(x)}\). In particular, we consider a new compact subset \(K'\) of \(K\), such that \(K'\in I^+(z)\) for some point \(z\in Z\). By achronality of \(Z\), \(K'\cap Z=\emptyset\). Clearly, then, \(K'(\subset K)\) is a compact set avoiding \(\partial I^+(x)\) and thus \(K'\subset M-\overline{I^+(x)}\). Thus, we may always take our compact set \(K\) to such that \(K\subset M- \overline{I^+(x)}\) and \(K\subset I^-(y') - [\overline{I^-(y)}\cup \overline{I^-(\Sigma)}]\). \\ \indent If \((M,g)\) is outer continuous at \(x\), then there is an open neighborhood \(O(x)\) of \(x\) such that for any \(x'\in O(x)\), \(\overline{I^+(x')}\cap K=\emptyset\). As remarked above, any open neighborhood of \(x\) contains a subneighborhood \(U(x)\) such that \(U(x)\cap I^-(\mathcal{J}^+)\neq \emptyset\). Let \(x'\) be any point in this intersection, i.e., \(x'\in U(x)\cap I^-(\mathcal{J}^+)\). By conditions on \(\gamma\), \(y\), \(y'\), for a sufficiently small subneighborhood \(U(x)\), any such \(x'\) lies in \(I^-(y')\). \\ \indent Now consider a point \(w\) in \(I^-(k) - K\) for some point \(k\in int(K)\). Choose \(w\) such that \(x'\in I^-(w)\). This choice is possible in virtue of the fact that \(K\subset I^-(y')\) and \(x'\in I^-(y')\). Now, if \(x'\in I^-(k)\), then \(I^-(x')\subset I^-(k)\). By reflectivity this implies that \(I^+(k)\subset I^+(x')\). Since \(K\cap I^+(k)\neq \emptyset\), we have \(K\cap I^+(x')\neq \emptyset\). So \(K\not\subset M - \overline{I^+(x')}\). Since \(U(x)\) can be chosen to be arbitrarily small and \(x'\) is an arbitrary point in \(U(x)\), \((M,g)\) is not outer continuous at \(x\). 
\end{proof}
The diagram arising in the above proof.  
\begin{center}
\includegraphics[width=.6\linewidth]{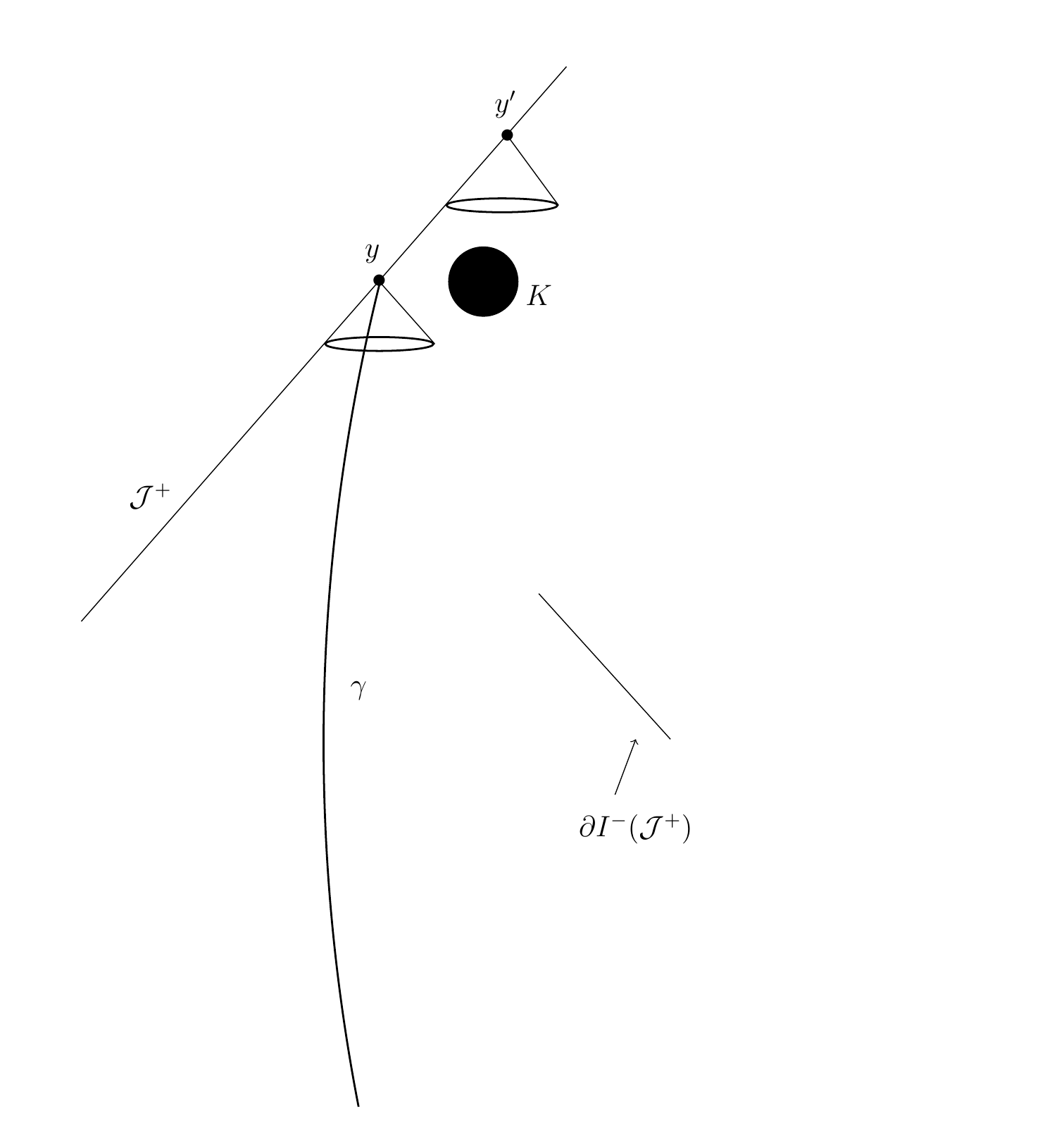}
\end{center}

\section{Discussion}
A. \textbf{On causal discontinuity}. \\ \indent Hawking and Sachs \cite{HS} formulated causal continuity in search of a property weaker than global hyperbolicity but stronger than stable causality. They argue that global hyperbolicity may be too strong for a number of physically interesting models, and that stable causality includes spacetimes with curious causal properties. They consider the example of Minkowski spacetime with a horizontal line removed. This example is stably causal, but it has the following pathological feature: the extent of the spacetime covered by \(I^-(p)\) changes discontinuously upon arbitrarily small perturbations of the metric. Their proposal of causal continuity is an effort to prohibit such pathologies. \\ \indent By theorem 4.1, if spacetimes containing fully evaporated black holes are causally continuous, then one or more of the assumptions must be violated. Which one of these assumptions is most vulnerable is a question for future investigations. The appeal of theorem 4.1 is that each condition is either entirely standard, or representative of some property which is almost assumed in models of evaporating black holes. Thus, we conclude that if Lorentzian manifolds provide faithful macroscopic models of evaporating black hole spacetimes, then either these are causally discontinuous, or there is a major departure from the common picture of evaporation. \\ \\
B. \textbf{On predictability}. \\ \indent Wald \cite{Wald} remarks that non-global hyperbolicity need not imply the failure of predictability. He considers the example of a massless Klein Gordon scalar field \(\phi\) in Minkowksi spacetime with a point removed, i.e., \((\mathbb{R}^4 \backslash{p},\eta)\). If we take \(S_1\) and \(S_2\) to lie, respectively, below and above the removed point, then \(S_2 \not\subset D^+(S_1)\), as in the conclusion of theorem 1.1 and 3.1. Yet Wald suggests that, given suitable initial data, \(\phi\) can still be determined globally and uniquely on \((\mathbb{R}^4 \backslash{p},\eta)\). Though his argument is heuristic, it seems that it could be made rigorous and that it could be generalized to other matter fields. On the basis of this example, Wald suggests that fully evaporating black hole spacetimes could be of this kind, i.e., non-globally hyperbolic yet still permitting unique global evolution of matter fields residing in the spacetime.\footnote{Private communication: Wald still stands by those arguments.} \\ \indent
The important point here is that Wald’s example \textit{is} causally continuous. This raises the question of whether his argument goes through in causally discontinuous spacetimes. On the basis of some simple examples constructed by removing lines in Minkowski spacetime, it seems not. Unfortunately, reasoning by examples is rarely enough. What we would like to have is a theorem (or counterexample) to the effect that causally discontinuous spacetimes prevent a system of hyperbolic PDEs from admitting globally unique solutions determined by initial data. Such a result would express, clearly and rigorously, once and for all, the widely held suspicion that full evaporation conflicts with predictability. \\ \\
C. \textbf{Holography and AdS/CFT}\\ \indent
The last two decades have seen much interest in ideas going under the general heading of AdS/CFT and holography. There is a widely discussed and popular idea to model evaporating black hole in spacetimes with asymptotic boundaries which are timelike (eg. AdS, Schwarzchild-AdS, etc). It has been widely suggested that in those spacetimes, full black hole evaporation does not lead to a failure of predictability.\footnote{Note that a mathematically rigorous underpinning of this idea is still lacking.} Here, we point out that theorem 4.1 combined with recent results in \cite{AM} have bearing on this particular issue. \\ \indent The first point is that it is clear that it is possible to formulate theorem 4.1 for both timelike and spacelike boundaries. Thus, a spacetime with timelike boundary containing a fully evaporating black hole is causally discontinuous.  \\ \indent In addition to this point, we mention the recent study \cite{AM}, which is a detailed investigation of the causal structure of spacetimes with timelike boundaries. These are spacetimes that are endowed with timelike boundaries, which we represent formally \(\overline{M}=M\cup \partial M\). It is a theorem that such spacetimes cannot be globally hyperbolic in the usual sense. The best possible causal property for \(\overline{M}\) is `globally-hyperbolic-with-timelike-boundary', which is defined analogously to the usual notion of global hyperbolicity. The authors of \cite{AM} prove various properties that are implied by being globally-hyperbolic-with-timelike-boundary. Combining theorem 4.1 with theorem 3.8 in \cite{AM}, we deduce that no spacetime \(M\) satisfying the conditions of theorem 4.1 can admit a timelike conformal boundary such that the spacetime with boundary \(\overline{M}\) is globally-hyperbolic-with-timelike-boundary. Note, however, that this is not true of theorem 3.1 for, as in remark 3.8(b) of \cite{AM}, \(M\) may be non-causally simple even if \(\overline{M}\) is globally hyperbolic with timelike boundary. \\ \indent 
This corollary of theorem 3.8 in \cite{AM} and theorem 4.1 bears on the prospect of predictability in the evaporating context. For, though failure of global hyperbolicity makes predictability less likely, failure of globally-hyperbolic-with-timelike-boundary is even worse. This has a consequence for the AdS/CFT and holography community, where it is often suggested that evaporation in spacetimes with timelike boundaries can provide for predictable evolution of fields. If correct then given the previous paragraph, either there are major departures from the global causal conditions of theorem 4.1, or, what is meant by `predictable' in the AdS/CFT or holography context is truly different from the notion of predictability proper to the global spacetime context. Further results in either direction are in high demand.

\newpage

\section*{Appendix A - Proof of theorem 1.1}
\begin{proof}
First, note that by by p.414 corollary 26 and also p.413, proposition 25 of \cite{ON}, that \(S_1\) and \(S_2\) are achronal and edgeless implies that these are also topological manifolds. This will be used below. \\ \indent 
Suppose, for contradiction, that \(S_2 \subset D^+(S_1)\). \\ \indent Then we can consider a smooth, non-vanishing timelike vector field \(t\) on \(M\). Since \(S_1\) and \(S_2\) are achronal, no integral curve of \(t\) can intersect either of these sets more than once. However, by \(S_2 \subset D^+(S_1)\), each integral curve which intersects \(S_2\) must intersect \(S_1\). Thus, by following these integral curves we obtain a one to one projection \(f\) from \(S_2\) onto a subset \(A\) of \(S_1\). Since \(t\) is smooth, and \(S_1\) and \(S_2\) are topological manifolds transverse to the \(t-\)flow, \(f: S_2\to A\) is a homeomorphism. Moreover, since \(S_2\) is edgeless, by the same argument as in proposition 6.3.1 of \cite{HE}, it is an imbedded \(C^0\) manifold (without boundary). This implies that \(A = f(S_2)\) must be an open subset of \(S_1\). However, we can prove that \(A\) is closed as follows. \\ \indent Let \(q \in \overline{A}\) and suppose \(q \not\in A\). Then, since \(q \not\in A\), we clearly have a future inextendible timelike curve from \(q\) which does not intersect \(S_2\), and hence \(q \not\in D^-(S_2)\) - see proposition 6.5.1 of \cite{HE}. This implies that \(q \in K\). On the other hand, since \(q \in \overline{A}\), there exists a sequence \(\{q_n\}\) in \(A\) which converges to \(q\). Let \(r_n = f^{-1}(q_n)\). Since infinitely many \(q_n\) enter \(int(K)\), infinitely many \(r_n\) lie in \(J^+(K) \cap S_2\). By compact closure, the sequence \(\{r_n\}\) must have an limit point \(r\) in \(S_2\). By continuity of \(f\), we must have \(f(r) = q\). This contradicts the assumption that \(q \not\in \overline{A}\) and thus proves that \(A\) is closed. \\ \indent Since \(S_1\) is connected, the fact that \(f(S_2)=A\) is both open and closed means that \(f(S_2)=A=S_1\). However, this is impossible since the integral curve of \(t\) which passes through the point \(p \in D^+(S_1)\), \(p \not\in J^+(S_2)\cup J^-(S_2)\) must intersect \(S_1\) at a point not lying in \(A\). Thus \(S_2 \not\subset D^+(S_1)\). 
\end{proof}

\end{document}